\documentclass[conference,letterpaper]{IEEEtran}

\usepackage{amsmath}
\usepackage{amssymb}
\usepackage{balance}

\usepackage{bm}
\usepackage{cite}
\usepackage{color}
\usepackage[T1]{fontenc}
\usepackage[cmintegrals]{newtxmath}
\usepackage{graphicx}
\usepackage[utf8]{inputenc}
\usepackage{longtable}
\usepackage{mathrsfs}
\usepackage{mathtools}
\usepackage{multirow}
\usepackage{float}
\usepackage{siunitx}
\usepackage{soul}               
\usepackage[color=green!40]{todonotes}
\soulregister\cite7
\soulregister\footnote7
\soulregister\eqref7
\soulregister\ref7
\usepackage{url}
\usepackage{threeparttable}
\def\BibTeX{{\rm B\kern-.05em{\sc i\kern-.025em b}\kern-.08em
    T\kern-.1667em\lower.7ex\hbox{E}\kern-.125emX}}
\graphicspath{{./figures/}}

\newcommand{\myauthor}{%
  \IEEEauthorblockN{Xintao Huan}
  \IEEEauthorblockA{\textit{Department of Electrical and Electronic Engineering} \\
    \textit{Xi'an Jiaotong-Liverpool University (XJTLU)}\\
    Suzhou, China \\
    Xintao.Huan@xjtlu.edu.cn}%
  \and%
  \IEEEauthorblockN{Kyeong Soo Kim}
  \IEEEauthorblockA{\textit{Department of Electrical and Electronic Engineering} \\
    \textit{Xi'an Jiaotong-Liverpool University (XJTLU)}\\
    Suzhou, China \\
    Kyeongsoo.Kim@xjtlu.edu.cn}%
  \thanks{This work was supported by Xi'an Jiaotong-Liverpool University
    Research Development Fund (RDF) under grant reference number RDF-16-02-39.}%
}

\newcommand{\mytitle}{Optimal Message Bundling with Delay and Synchronization
  Constraints in Wireless Sensor Networks}%

\IEEEoverridecommandlockouts    

\begin{document}

\title{\mytitle}

%

\author{\myauthor}

\maketitle

\begin{abstract}
  Message bundling is an effective way to reduce the energy consumption for
  message transmissions in wireless sensor networks. However, bundling more
  messages could increase both end-to-end delay and message transmission
  interval; the former needs to be maintained within a certain value for
  time-sensitive applications like environmental monitoring, while the latter
  affects time synchronization accuracy when the bundling includes
  synchronization messages as well.
  Taking as an example a novel time synchronization scheme recently proposed for
  energy efficiency, we propose an optimal message bundling approach to reduce
  the message transmissions while maintaining the user-defined requirements on
  end-to-end delay and time synchronization accuracy. Through translating the
  objective of joint maintenance to an integer linear programming problem, we
  compute a set of optimal bundling numbers for the sensor nodes to constrain
  their link-level delays, thereby achieve and maintain the required end-to-end
  delay and synchronization accuracy while the message transmission is
  minimized.
\end{abstract}


\IEEEpeerreviewmaketitle

\section{Introduction}
\label{sec:introduction}
\begin{figure}[!tb]
  \begin{center}
    \includegraphics[width=\linewidth]{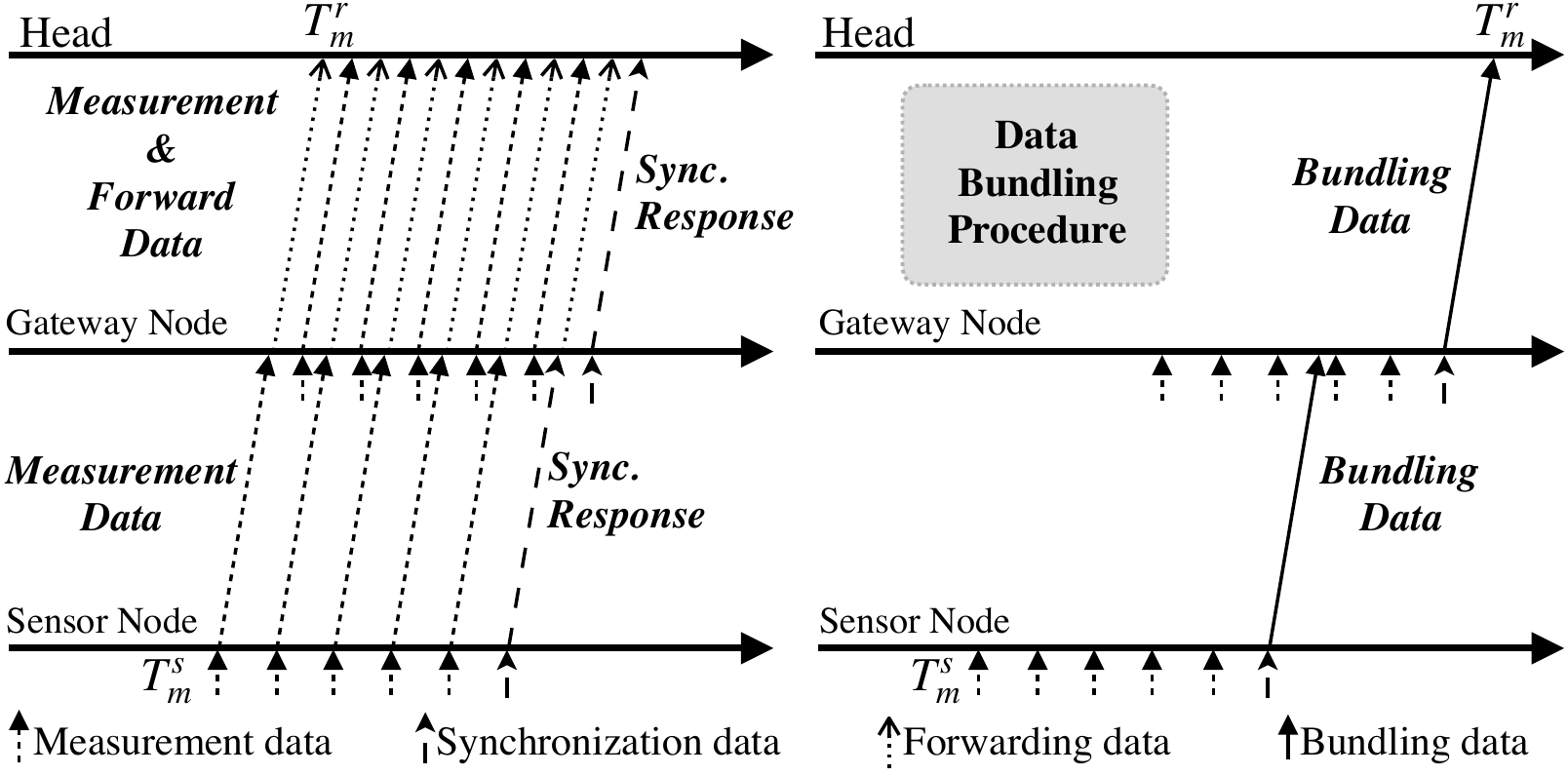}
  \end{center}
  \caption{Comparison of the message transmissions of the time synchronization
    schemes with and without data bundling procedure.}
  \label{fig:data_bundling}
\end{figure}

In the typical wireless sensor networks (WSNs), with a view to the energy
capacity and the serve time, minimizing energy consumption is utmost critical.
Considering the radio activities consume the most energy in the sensor node,
reducing the number of message transmissions in the sensor network is considered
as a major approach of energy conservation, in which an efficient way is the
data bundling\footnote{The terms of “data bundling” and “message bundling” are 
used interchangeably in this paper.} \cite{DB:survey}. Depending on the
operation locations, the existing data bundling schemes
(e.g.,\cite{DB:con1,DB:con2,DB:con3}) could be classified to in-node, in-network
and hybrid bundling. However, one common negative impact of employing the data
bundling procedure is the increasing of end-to-end (E2E) delay. By E2E delay, we
mean the difference between the time of measurement ($T_{m}^s$) at an
originating sensor node and the time of the reception of the resulting
measurement data by the head node ($T_{m}^r$) via a message as demonstrated in
Fig.~\ref{fig:data_bundling}.
Nevertheless, for most monitoring applications in WSNs, time synchronization is
also critical for ordering the measurement data and detecting the
events. Consequently, optimally bundling the transmission messages with
maintaining the E2E delay and the time synchronization accuracy is in demand,
which is not considered in the existing data bundling schemes.

On the basis of the novel energy-efficient time synchronization scheme
\cite{Kim:17-1}---called EE-ASCFR throughout the paper---which proposes data
bundling as an optional procedure for reducing energy consumption, the time
synchronization could be realized without heavy consumption on the energy and
computing resources.
In EE-ASCFR, because synchronization messages are embedded in measurement data
report messages from sensor nodes to reduce the number of message transmissions,
the synchronization interval (SI) is tied to the interval of report messages for
measurement data. Bundling more data, therefore, could reduce more energy consumption
for message transmissions, but it could not only increase E2E delay but also
worsen synchronization accuracy due to the increased SI as discussed in
\cite{Kim:17-1}. In such a case, we need to optimize the number of bundled
messages under the constraint of E2E delay and time synchronization accuracy in
achieving higher energy efficiency.

In this paper, we formulate the aforementioned bundling optimization problem as
integer linear programming (ILP),
where the number of bundled messages for each sensor node is optimized while
jointly satisfying the user-defined performance requirements on E2E delay and
time synchronization accuracy. In this way, we can further reduce the energy
consumption through data bundling while meeting the requirements.

The rest of the paper is organized as follows: Section~\ref{sec:preliminaries}
introduces the time synchronization scheme based on the data bundling procedure
as well as the related conflicts of performance
metrics. Section~\ref{sec:our_approach} presents our proposed approach and its
formulation as ILP. Section~\ref{sec:system_design} exhibits our system design
at both head\footnote{The head node and the PC or server connected to it are 
jointly called as head throughout this paper.} and sensor
node. Section~\ref{sec:experimental_evaluation} demonstrates the performance of the
proposed approach through experimental results on a real WSN
testbed. Section~\ref{sec:concluding_remarks} concludes our work in this paper
and discusses the future works.

\section{Preliminaries}
\label{sec:preliminaries}
Most WSNs have the fundamental characteristics of limited resources, multi-hop
communication, large scale and dynamic environments \cite{WSN:charas}. To
achieve satisfactory performance in typical WSN applications including
environmental monitoring, event detection and industrial applications, several
specific requirements---e.g., E2E delay and synchronization accuracy---have to
be met.
With consideration to the most basic requirement of energy efficiency, joint
maintaining the three metrics are crucial.

\subsection{Energy-Efficient Time Synchronization Schemes Using Data Bundling}
\label{sec:time_synchronization}
\begin{figure}[!tb]
  \begin{center}
    \includegraphics[width=\linewidth]{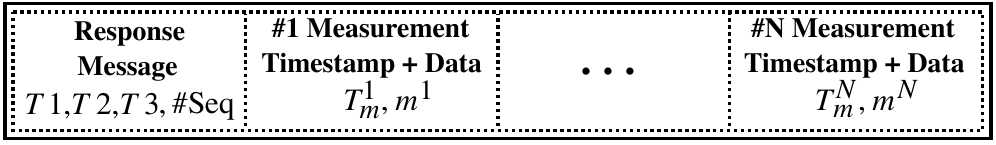}
  \end{center}
  \caption{Payload contents of the data bundling procedure introduced in EE-ASCFR \cite{Kim:17-1}.}
  \label{fig:payload}
\end{figure}
\begin{table}[!tb]
  \centering
  \begin{threeparttable}
    \caption{MAE and MSE of Measurement Time Estimation of EE-ASCFR and AHTS
      with Different SIs Provided in \cite{Kim:17-1} and \cite{Huan:19-1}}
    \label{tab:time_sync_results_si}
    \centering \setlength{\tabcolsep}{4mm}
   	\begin{tabular}{|c|l||r|r|}
      \hline
      \multicolumn{2}{|c||}{Synchronization Scheme}
      & \multicolumn{1}{c|}{MAE \tnote{1}}
      & \multicolumn{1}{c|}{MSE \tnote{2}} \\ \hline\hline
      \multirow{3}{*}{EE-ASCFR}
      & SI$\;=100~\mbox{s}$ & 8.8811E-25 & 5.8990E-19 \\ \cline{2-4}
      & SI$\;=1~\mbox{s}$ & 9.1748E-25 & 5.4210E-19 \\ \cline{2-4}
      & SI$\;=10~\mbox{ms}$ & 1.0887E-24 & 4.7684E-19 \\ \hline\hline
      \multirow{3}{*}{AHTS}
      & SI$\;=100~\mbox{s}$  & 8.4225E-06 & 1.2524E-10 \\ \cline{2-4}
      & SI$\;=10~\mbox{s}$  & 2.3385E-06 & 9.1694E-12  \\ \cline{2-4}
      & SI$\;=1~\mbox{s}$  & 1.8166E-06 & 5.2094E-12  \\ \hline
    \end{tabular}
    \begin{tablenotes}
    \item[1] MAE is the mean absolute error of measurement time estimation.
	 \item[2] MSE is the mean square error of measurement time estimation.
    \end{tablenotes}
  \end{threeparttable}
\end{table}
To fulfill the desired requirements of energy efficiency and synchronization
accuracy, many schemes have been proposed such as
\cite{Kim:17-1,EETS,CESP}. Among those, EE-ASCFR proposed in \cite{Kim:17-1}
particularly suits the E2E delay calculation since computing the E2E delay at
the head is completely in conformity with the preferential asymmetric scenario
of EE-ASCFR. However, because the synchronization accuracy (SA) is affected by
SI as illustrated in Table~\ref{tab:time_sync_results_si} (i.e., simulation
results of EE-ASCFR and practical evaluation results of AHTS\cite{Huan:19-1}),
we need to maintain SI to a reasonable value for better synchronization
accuracy. Based on those evaluation results in
Table~\ref{tab:time_sync_results_si}, the relationship ($\mathscr{T}$) between
SI and synchronization accuracy is represented as follows:
\begin{equation}
  \label{eq:relationship}
  SI = \mathscr{T}(SA),
\end{equation}
in which the requirement of SA could be translated to the requirement of SI.

To further reduce energy consumption, EE-ASCFR also employs data bundling: In
downstream, the synchronization ``Request'' message for timestamp $T1$ is
embedded inside a regular beacon message; in upstream, the synchronization
``Response'' message for timestamps $T1$, $T2$, and $T3$ are bundled together
with measurement data as shown in Fig.~\ref{fig:payload}. Note that bundling
more measurement data could lead to better energy efficiency, which, however,
increases SI and decreases SA as discovered in the evaluation of \cite{Kim:17-1}
and \cite{Huan:19-1}. Nevertheless, as EE-ASCFR only covers the single-hop
scenario, the more asymmetric approach (i.e., AHTS), which is based on EE-ASCFR
and extends it to multi-hop resource-constrained sensor networks, is employed in
the proposed approach to cover multi-hop scenarios. The data bundling procedure
is inherited and extended to the multi-hop case as illustrated in
Fig.~\ref{fig:data_bundling}.


\subsection{Conflicts of Performance Metrics: Energy Efficiency, E2E Delay and
  Synchronization Accuracy}
\label{sec:req_conflicts}
%
%
Simultaneously meeting the requirements for the three performance metrics of
energy efficiency, E2E delay, and synchronization accuracy is not possible due
to their relationship discussed in
Section~\ref{sec:time_synchronization}. Time synchronization provides the
possibility of accurate calculation and maintenance of E2E delay, and employing
the time synchronization scheme requires certain computing and power resources,
which turns out the \textit{conflict between energy efficiency and E2E delay 
calculation}. Although the data bundling procedure introduced in EE-ASCFR
could drastically reduce the energy consumption for message transmissions and
bundling more data could lead to higher energy efficiency, the data bundling
procedure could direct result in high E2E delay as exhibited in
Fig.~\ref{fig:data_bundling}, which is the \textit{conflict between energy 
efficiency and E2E delay maintenance}. Specific to the delay calculation
illustrated in Fig.~\ref{fig:data_bundling}, the E2E delay---i.e., $T_{m}^r$ -
$T_{m}^s$---of measurement $m$, would be relatively small (e.g., multiples of
forwarding delay which typically in milliseconds) in the regular direct
forwarding method. However, due to the bundling procedure in the intermediate
sensor nodes, the E2E delay could be as large as multiples of measurement
interval. Nonetheless, the bundling procedure is shown to be quite efficient in
conserving energy by reducing the number of message transmissions as shown in
Fig.~\ref{fig:data_bundling}, which is critical to low-power sensor nodes. As
for synchronization accuracy, SI should be shorter for achieving higher accuracy 
\cite{Kim:17-1}, but shorter SI results in more message transmissions, which 
again leads to higher energy consumption. This draws forth the \textit{conflict 
between energy efficiency and synchronization accuracy}. To jointly meet the three
performance requirements, comprehensive optimizations should be taken. To yield
comprehensive satisfactory performances among those three requirements, a new
approach for the bundling number optimization for each sensor node is
proposed to fulfill the requirement of maintaining E2E delay and
synchronization accuracy while minimizing the message transmissions.

\section{ILP Model for Optimal Bundling Problem}
\label{sec:our_approach}
In a WSN with $N$ sensor nodes, the energy consumption ${e}_{i}^{t}$ for the
message transmissions at sensor node $i$ is modeled as follows: For
$i{\in}\left[0,1,\ldots,N{-}1\right]$,
\begin{equation}
  \label{eq:ec_original}
  {e}_{i}^{t} = \alpha_{i}{e}_{i}^{m} + \beta_{i}{e}_{i}^{s} + \gamma_{i}{e}_{i}^{f}, 
\end{equation}
where ${e}_{i}^{m}$ and ${e}_{i}^{s}$ denote the energy consumption for the
transmission of a measurement and a synchronization message generated by sensor
node $i$ respectively, and ${e}_{i}^{f}$ is the energy consumption for
forwarding either a synchronization or a measurement message from offspring
sensor nodes at sensor node $i$. The coefficients $\alpha_{i}$, $\beta_{i}$ and
$\gamma_{i}$ are the number of transmissions for corresponding messages.

If we apply data bundling, \eqref{eq:ec_original} can be modified as follows:
\begin{equation}
  \label{eq:ec_bundling}
  {e}_{i}^{t} = 
  \begin{cases}
		\delta_{i}{e}_{i}^{b},& \text{all data bundling}\\
		\delta_{i}{e}_{i}^{b} + \gamma_{i}{e}_{i}^{f},& \text{self data bundling}
	\end{cases}
\end{equation}
where $\delta_{i}{e}_{i}^{b}$ is the total energy consumption caused by
transmitting the bundled messages. Note that, unlike the all data bundling
option which bundles all data into one bundled message, the self data bundling
option bundles only the data generated from the sensor node itself, which is the
reason the term $\gamma_{i}{e}_{i}^{f}$ still remains. From the network-level
perspective, the total energy consumption for message transmissions in the
network could be described as follows:
\begin{equation}
  \label{eq:ec_network}
  \mathbf{E}^{t} = \sum_{i=0}^{N-1} {e}_{i}^{t}.
\end{equation}

\subsection{Maximization of Bundling Number for Energy Efficiency}
\label{sec:maxim-bundl-numb}
With the model for energy consumption caused by message transmissions, we can
increase the network energy efficiency by minimizing the total energy
consumption for message transmissions (i.e., $\mathbf{E}^{t}$).

As shown in \eqref{eq:ec_bundling}, a large number of bundled message
transmissions (i.e., $\delta_{i}$) could result in more energy consumption, but
it could be decreased by bundling more messages in one transmission. Let
$\Gamma^{i}$ be the number of bundled messages at sensor node $i$. Then, a
larger $\Gamma^{i}$ could lead to a smaller $\delta_{i}$, which could reduce
total energy consumption for message transmissions. Consequently, we can
minimize $\mathbf{E}^{t}$ by maximizing $\Gamma^{i}$.

Even though we can increase the number of bundling for better energy efficiency,
we cannot indefinitely because the E2E delay of the measurement data is
sensitive to the number of bundling. Therefore, the maximization of bundling
number for energy efficiency can be formulated as follows:
\begin{equation}
  \label{eq:optimal_bundling}
  \begin{split}
    & \mathbf{maximize} ~ \mathbf{\Gamma} \\
    & \mathbf{subject~to} ~ \chi^{min} \leq \Gamma^{i} \leq \chi^{max}, \forall
    i \in [0,\ldots,N{-}1],
  \end{split}
\end{equation}
where
\[
  \mathbf{\Gamma} = \sum_{i=0}^{N-1} {\Gamma}^{i}.
\]
Note that $\mathbf{\Gamma}$ is the total bundling number in the network and that
$\chi^{min}$ and $\chi^{max}$ are the lower and upper bounds of the measurement
bundling number which are application-specific parameters and could be specified
by user.

%
%
%
%
%
%
%
%
%

\subsection{Constraining E2E Delay}
\label{sec:constraining_e2e_delay}
We define the E2E delay of sensor node $i$ as the difference between the time of
a certain measurement $m$ at the sensor node and the time of the reception of
the resulting message by the head:
\begin{equation}
  \label{eq:e2e_delay}
  \mathbf{D}_{e2e}^{i} \triangleq \sum_{l=0}^{L-1} {D}_{l}^{i} = T_{m}^{i,r} - T_{m}^{i,s},
\end{equation}
where ${D}_{l}^{i}$ is the link delay at link $l$ of $L$ links from sensor node
$i$ to the head, $T_{m}^{i,r}$ and $T_{m}^{i,s}$ are the receiving time at the
head and the measuring time at the sensor node respectively, the latter of which
is a time with respect to the reference clock at the head translated by a time
synchronization scheme.
Specifically, the $\mathbf{D}_{e2e}^{i}$ is a path-level delay which consists of
several link-level delays in the network. Considering the bundling procedure,
the link delay at link $l$ for sensor node $i$ could be described as follows:
\begin{equation}
  \label{eq:link_delay}
  D_{l}^{i} = D_{prop}^{i,l} + D_{serv}^{i,l} + D_{bund}^{i,l},
\end{equation}
where $D_{prop}^{i,l}$ denotes the propagation delay which is typically in
nanosecond level in WSN, $D_{bund}^{i,l}$ is the delay caused by the bundling
procedure which is in multiples of the measurement interval (e.g.,
$5 \times $\SI{1}{\s} for the measurement interval of \SI{1}{\s} and the
bundling number of 5). Based on the service time model for TinyOS
\cite{Fu:ICDCS}, we can model $D_{serv}^{i,l}$ as follow:
\begin{equation}
  \label{eq:service_delay}
  D_{serv}^{i,l} = 
  \begin{cases}
    D_{SPI}^{i,l} + D_{succ}^{i,l} + (N_{try}^{i,l}-1)\cdot D_{retry}^{i,l},& \text{$N_{try}^{i,l}\leq N_{max}^{i,l}$}\\
    D_{SPI}^{i,l} + D_{fail}^{i,l} + (N_{max}^{i,l}-1)\cdot D_{retry}^{i,l},&
    \text{$N_{try}^{i,l} > N_{max}^{i,l}$}
  \end{cases}
\end{equation}
where
\[
\begin{split}
D_{succ}^{i,l} & = D_{MAC}^{i,l} + D_{frame}^{i,l} + D_{ACK}^{i,l}, \\
D_{fail}^{i,l} & = D_{MAC}^{i,l} + D_{frame}^{i,l} + D_{waitACK}^{i,l}, \\
D_{retry}^{i,l} & = T_{retry}^{i,l} + D_{frame}^{i,l} + D_{waitACK}^{i,l}.
\end{split}
\]
Note that, the delay parameters---i.e., one-time serial-peripheral interface
(SPI) bus loading delay $D_{spi}^{i,l}$, medium access control (MAC) layer delay
$D_{MAC}^{i,l}$, frame transmission delay $D_{frame}^{i,l}$, acknowledgment
(ACK) transmission delay $D_{ACK}^{i,l}$ and ACK waiting delay
$D_{waitACK}^{i,l}$---in the above equations are platform-dependent values, and
their values are typically in the order of milliseconds. In addition,
$N_{try}^{i,l}$ and $N_{max}^{i,l}$ are the current and the maximum allowed
number of transmissions for a successful delivery, and $T_{retry}^{i,l}$ is the
user-defined backup time of retransmission. For simplicity and energy
efficiency, the packet retransmission is not taken into account in our proposed
scheme since there are usually not many packet retransmissions in the network
with lower traffic.
So \eqref{eq:service_delay} could be simplified as follows:
\begin{equation}
  \label{eq:service_delay_sp}
  D_{serv}^{i,l} = D_{SPI}^{i,l} + D_{MAC}^{i,l} + D_{frame}^{i,l} + D_{ACK}^{i,l}
\end{equation}
where the value of $D_{serv}^{i,l}$ is around \SI{10}{\ms} based on the
reference values in \cite{Fu:ICDCS}.

The link delay in \eqref{eq:link_delay} could be further simplified when the
measurement interval for an application ($\mathbf{I}_{meas}^{i,l}$) is much
larger than the service delay ($D_{serv}^{i,l}$): Because
$D_{bund}^{i,l}{>}\mathbf{I}_{meas}^{i,l}$,
$\mathbf{I}_{meas}^{i,l}{\gg}D_{serv}^{i,l}$ also implies
$D_{bund}^{i,l}{\gg}D_{serv}^{i,l}$. Hence
\begin{equation}
  \label{eq:link_delay_new}
  D_{link}^{i,l} \approx D_{bund}^{i,l}, ~ \mbox{ if } \mathbf{I}_{meas}^{i,l} \gg D_{serv}^{i,l}.
\end{equation}
Note that the service delay $D_{serv}^{i,l}$ should not be ignored in case the
application requires frequent measurements, i.e., $\mathbf{I}_{meas}^{i,l}$ is
comparable to $D_{serv}^{i,l}$.

In typical hierarchical multi-hop WSNs, sensor nodes located in different layers
handle different amount of traffic: For instance, the gateway node in the upper
layer has to handle the message traffic from its offspring sensor nodes as well
as itself; the higher layer it is located in, the more message traffic it has to
handle. This means that even two sensor nodes with the same bundling number
(i.e., $\Gamma^{i}$) could have different bundling delays due to the variance in
their message traffics. By introducing a message traffic coefficient
($\frac{1}{1+\lambda^{i}}$) for each sensor node that periodically measures data
with the same measurement interval (i.e., $\mathbf{I}_{meas}^{i}$), the bundling
delay at sensor node $i$ ($D_{bund}^{i}$) could be represented as follows:
\begin{equation}
  \label{eq:bundling_delay}
  D_{bund}^{i} = \frac{\Gamma^{i}}{1+\lambda^{i}} \cdot \mathbf{I}_{meas}^{i},
\end{equation}
where $\lambda^{i}$ denotes the number of offspring sensor nodes. Then the
$D_{e2e}^{i}$ for the applications with normal measurement interval could be
modeled as follows:
\begin{equation}
  \label{eq:e2e_delay_new}
  \mathbf{D}_{e2e}^{i} = \sum_{l=0}^{L-1} D_{bund}^{i}.
\end{equation}
With \eqref{eq:e2e_delay_new}, we can also constrain the E2E delay in the
optimal bundling problem in \eqref{eq:optimal_bundling} with the user-defined
E2E delay requirement ($\mathbf{D}_{e2e}^{max}$): For
$i{\in}[0,1,\ldots,N{-}1]$,
\begin{equation}
  \label{eq:e2e_delay_constraints}
  \mathbf{D}_{e2e}^{i} \leq \mathbf{D}_{e2e}^{max}.
\end{equation}

\subsection{Constraining Synchronization Accuracy}
\label{sec:constraining_synacc}
Since the proposed optimal message bundling approach is based on the reverse asymmetric time
synchronization scheme, the synchronization accuracy depends on the report
interval of the synchronization messages which, in turn, are carried by the
bundled messages. In such a case, the user-required synchronization accuracy
($\mathbf{SA}^{min}$) could be maintained through constraining the E2E delay of
the bundled message when sensor nodes generate measurement data periodically.
Based on the empirical sets---i.e.,$\mathscr{T}$---of the relationships
between synchronization accuracy and SI previously provided in
Section~\ref{sec:preliminaries}, the user-required synchronization accuracy
could be translated to the delay requirement as follows:
\begin{equation}
\label{eq:synacc_translation}
  \mathbf{D}_{e2e}^{SA} = \mathscr{T}(\mathbf{SA}^{min}).
\end{equation}
Combining \eqref{eq:e2e_delay_new} and \eqref{eq:synacc_translation}, the
synchronization accuracy could be achieved through constraining the E2E delay as
follows: For $i{\in}[0,1,\ldots,N{-}1]$,
\begin{equation}
  \label{eq:synacc_constraints}
  \mathbf{D}_{e2e}^{i} \leq \mathbf{D}_{e2e}^{SA}.
\end{equation}

\subsection{ILP model}
\label{sec:ilp_model}
Combining the objective function~\eqref{eq:optimal_bundling} and the two
constraint sets \eqref{eq:e2e_delay_constraints} and
\eqref{eq:synacc_constraints}, we can formulate the optimal bundling problem as
the following ILP:
\begin{equation}
  \label{eq:ILP}
  \begin{split}
    & \mathbf{maximize} ~ \mathbf{\Gamma} = \sum_{i=0}^{N-1} {\Gamma}^{i} \\
    & \mathbf{subject~to~} \\
    &
    \begin{split}
      & ~~~~ \chi^{min} \leq \Gamma^{i} \leq \chi^{max}, ~
      \forall i \in [0,\ldots,N{-}1],\\
      & ~~~~ \mathbf{D}_{e2e}^{i} \leq \min\left(\mathbf{D}_{e2e}^{max},
        \mathscr{T}(\mathbf{SA}^{min})\right), ~ \forall i \in [0,\ldots,N{-}1],
    \end{split}
  \end{split}
\end{equation}
where the E2E delays of all sensor nodes are jointly constrained by the
user-defined E2E delay $\mathbf{D}_{e2e}^{max}$ and the synchronization accuracy
$\mathbf{SA}^{min}$ requirements, and the bundling number of each sensor node is
constrained by the user-defined lower $\chi^{min}$ and upper $\chi^{max}$
bounds, respectively. Applying the set of optimal bundling numbers computed from
this ILP model to the sensor nodes, the required E2E delay and synchronization
accuracy could be jointly maintained while the bundled message transmissions are
still minimized.

\section{System Design}
\label{sec:system_design}
\begin{figure}[!tb]
  \begin{center}
    \includegraphics[width=0.9\linewidth]{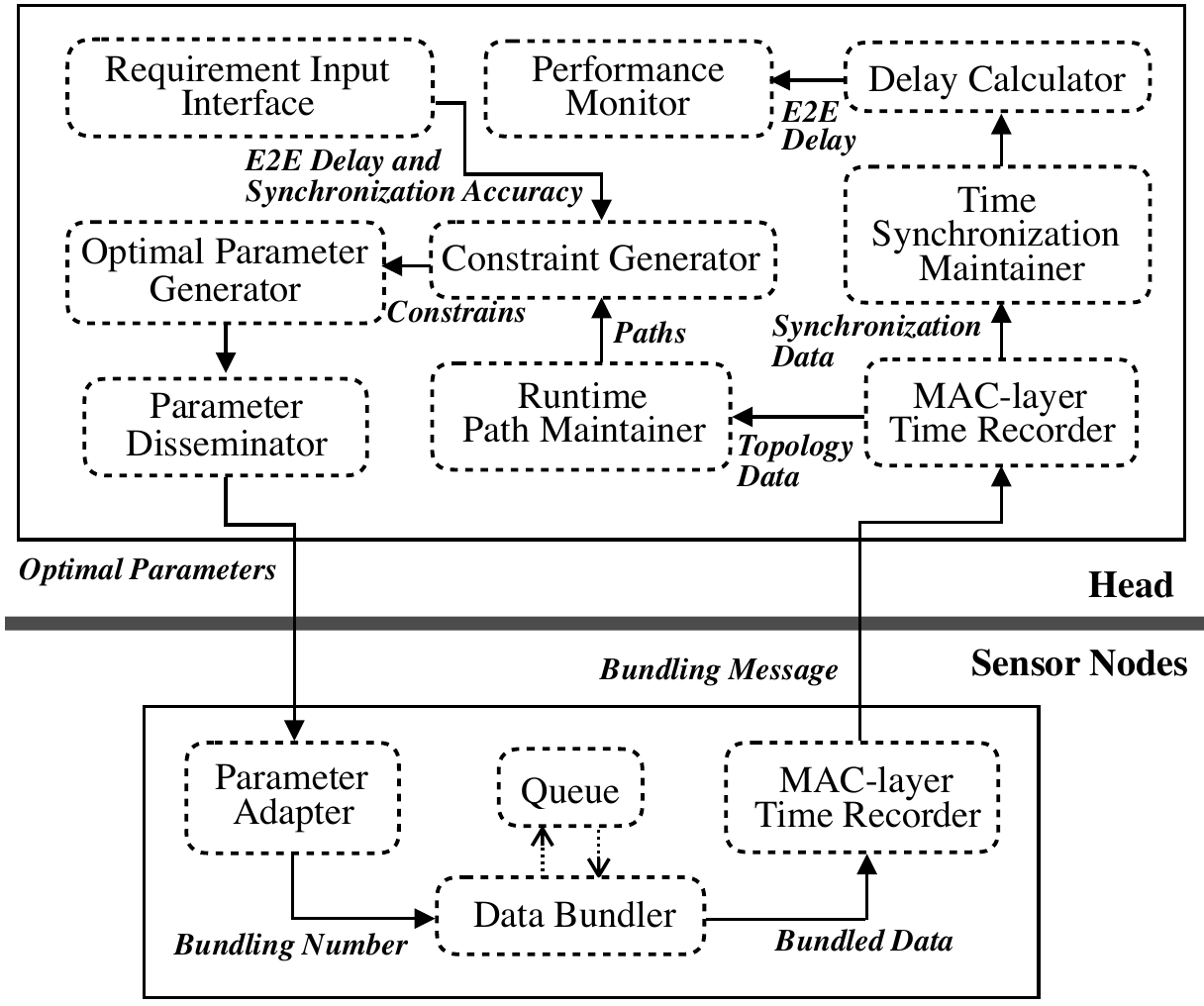}
  \end{center}
  \caption{System architecture of the proposed optimal bundling.}
  \label{fig:sys_archi}
\end{figure}

Fig.~\ref{fig:sys_archi} shows a system architecture for the proposed optimal
bundling based on the ILP model formulated in Section~\ref{sec:ilp_model}, where
the two subsystems---i.e., the \textit{performance maintainer} at the head and
the \textit{parameter adapter} at each sensor node---are built to achieve the
optimization target.

\subsection{Performance Maintainer at Head}
\label{sec:performance-maintainer}
Since the time synchronization operation is centralized at the head in EE-ASCFR
and AHTS, we first build the reference time synchronization system at the head
as a component of the proposed system. The time synchronization is achieved
through the \textit{MAC-layer Time Recorder} and the \textit{Time 
Synchronization Maintainer}, the latter of which translates timestamps between
the head and a sensor node. With the time synchronization component, a
measurement timestamp recorded at the sensor node---i.e., $T_{m}^{s}$ in
Fig.~\ref{fig:data_bundling}---is translated into a timestamp based on the
hardware clock of the head. Then, the E2E delay is obtained as a difference
between the translated measurement timestamp and the receiving timestamp of
$T_{m}^{r}$ through the \textit{Delay Calculator}. Afterwards, the runtime E2E
delay is monitored through the \textit{Performance Monitor}.

Based on the topology data carried in each bundled message, the routing paths
for all sensor nodes are recovered in the \textit{Runtime Path Maintainer}, and
one set of paths is generated and delivered to the \textit{Constraint 
Generator}. The user-defined requirements of E2E delay and synchronization
accuracy are captured through the user interface of \textit{Requirement Input 
Interface}. By combining the performance requirements and the current
path information, the \textit{Constraint Generator} generates a set of
constraints and passes it to the \textit{Optimal Parameter Generator}, where the
optimal bundling number for each sensor node is obtained as a solution of the
ILP model. Finally, the optimal bundling numbers from the \textit{Optimal 
Parameter Generator} are delivered to sensor nodes by the \textit{Parameter 
Disseminator}.

\subsection{Parameter Adapter at Sensor Nodes}
\label{sec:parameter-adapter}
To reduce the computational complexity required by the proposed optimal bundling on 
the sensor nodes, three lightweight components (including the \textit{MAC-layer Time Recorder}
from the time synchronization scheme) are implemented at sensor nodes. The
\textit{Parameter Adapter} receives the optimal bundling number and delivers it
to the \textit{Data Bundler} which bundles that number of measurement data
temporarily stored in the \textit{Queue} plus timestamps into one message. Then
the bundled message will be again timestamped for $T3$ by the \textit{MAC-layer
  Time Recorder} as shown in Fig.~\ref{fig:data_bundling}.

\section{Experimental Evaluation}
\label{sec:experimental_evaluation}

\begin{figure}[!tb]
  \begin{center}
    \includegraphics[width=\linewidth]{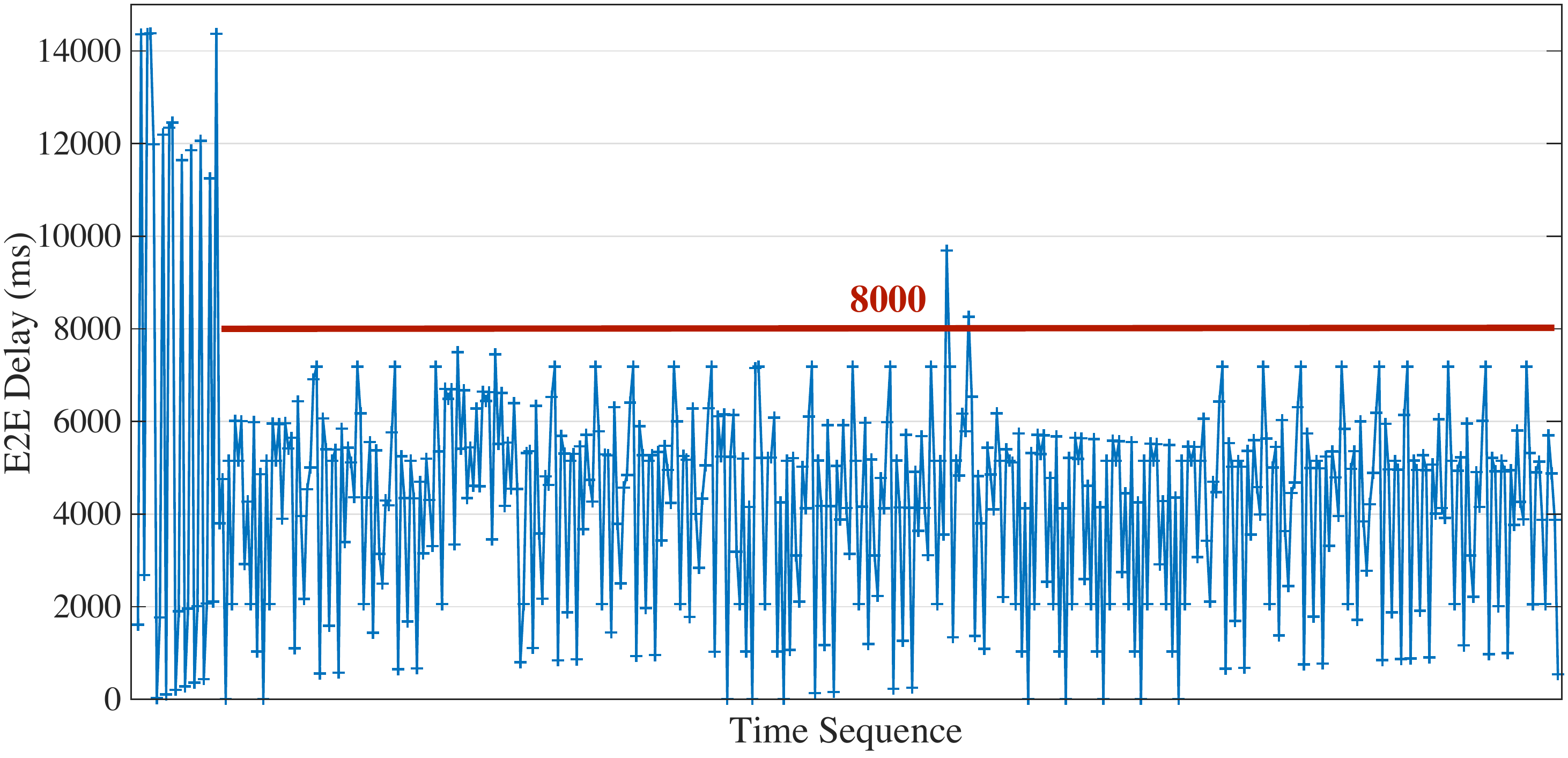}
  \end{center}
  \caption{E2E delay performance of the optimal bundling under static E2E delay
    requirement setting and the maximum bundling number of 15.}
  \label{fig:evaluation_onetime}
\end{figure}

\begin{figure}[!tb]
  \begin{center}
    \includegraphics[width=\linewidth]{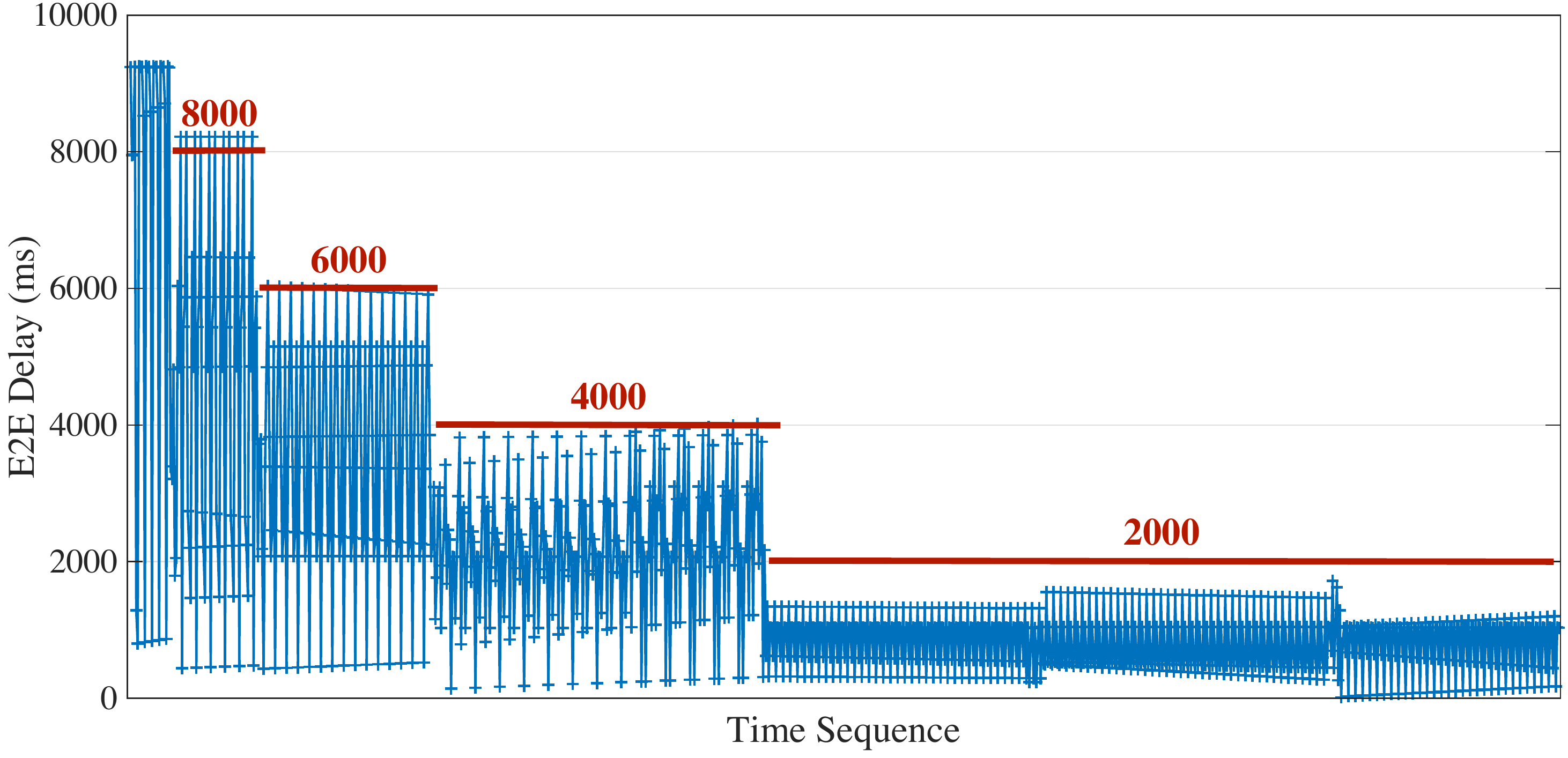}
  \end{center}
  \caption{E2E delay performance of the optimal bundling under dynamic E2E delay
    requirement setting and the maximum bundling number of 10.}
  \label{fig:evaluation_multi}
\end{figure}

The proposed approach is implemented on a real three-hop WSN testbed consisting
of five TelosB \cite{telosb} sensor nodes (i.e.,
head\,(\#0)${\leftrightarrow}$gateway\,(\#1)${\leftrightarrow}$gateway\,(\#2)${\leftrightarrow}$leaf\,(\#4) and 
head\,(\#0)${\leftrightarrow}$gateway\,(\#1)${\leftrightarrow}$leaf\,(\#3)).
During the experiments, each sensor node generates one measurement per second,
and the E2E delays of the latest measurements in bundled messages from all
sensor nodes are collected and stored in a time sequence in order of their
arrivals at the head, which are shown in Fig.~\ref{fig:evaluation_onetime} and
Fig.~\ref{fig:evaluation_multi}. The red horizontal lines with numbers (in
milliseconds) indicate the E2E delay requirements for corresponding time
periods, and the requirement of synchronization accuracy is set to \SI{5}{\us}
for all experiments.


\subsection{Delay Performance Under Static Requirement Setting}
\label{sec:evaluation_onetime}
We first evaluate the E2E delay performance of the optimal bundling under static
requirement setting and the maximum bundling number of $15$. We run the
experiment for \SI{3600}{\second} to demonstrate the \textit{long-term maintenance capability}.

As shown in the Fig.~\ref{fig:evaluation_onetime}, the E2E delays can exceed
\SI{14}{\s} before the optimal bundling is applied. Once the optimal bundling
is applied with the requirement of \SI{8}{\s}, however, we can see that the E2E
delay is controlled and kept under the requirement for most of the time. Note
that there are few data points crossed the requirement line; because the gateway
node serves its own measurement data first, the data from its offspring nodes,
sometimes, could be buffered in the queue and sent by the next available
message, which would increase the E2E delay of the corresponding message.

\subsection{Delay Performance Under Dynamic Requirement Setting}
\label{sec:evaluation_multi}
We also evaluate the E2E delay performance of the optimal bundling under dynamic
requirement setting with the maximum bundling number of $10$ to demonstrate its
\textit{run-time maintenance capability}. During the evaluation, the E2E delay
requirement is dynamically changed from \SI{8}{\s} to \SI{2}{\s} step-by-step as
shown in Fig.~\ref{fig:evaluation_multi}.

The results demonstrate that the proposed optimal bundling nicely handles
multiple requirements of E2E delay throughout the experiment. As discussed in
Section~\ref{sec:evaluation_onetime}, however, some data points slightly cross
the requirement line of \SI{8}{\s}, which is due to the neglect of the service
time (i.e., $D_{serv}^{i,l}$) in equation~\eqref{eq:link_delay}. When the
optimal bundling numbers are too strict which just fulfills the requirement, the
influence of the neglect of the small service time would be notable.

\subsection{Synchronization Accuracy and Energy Efficiency}
\label{sec:evaluation_syn}
\begin{figure}[!tb]
  \begin{center}
    \includegraphics[width=\linewidth]{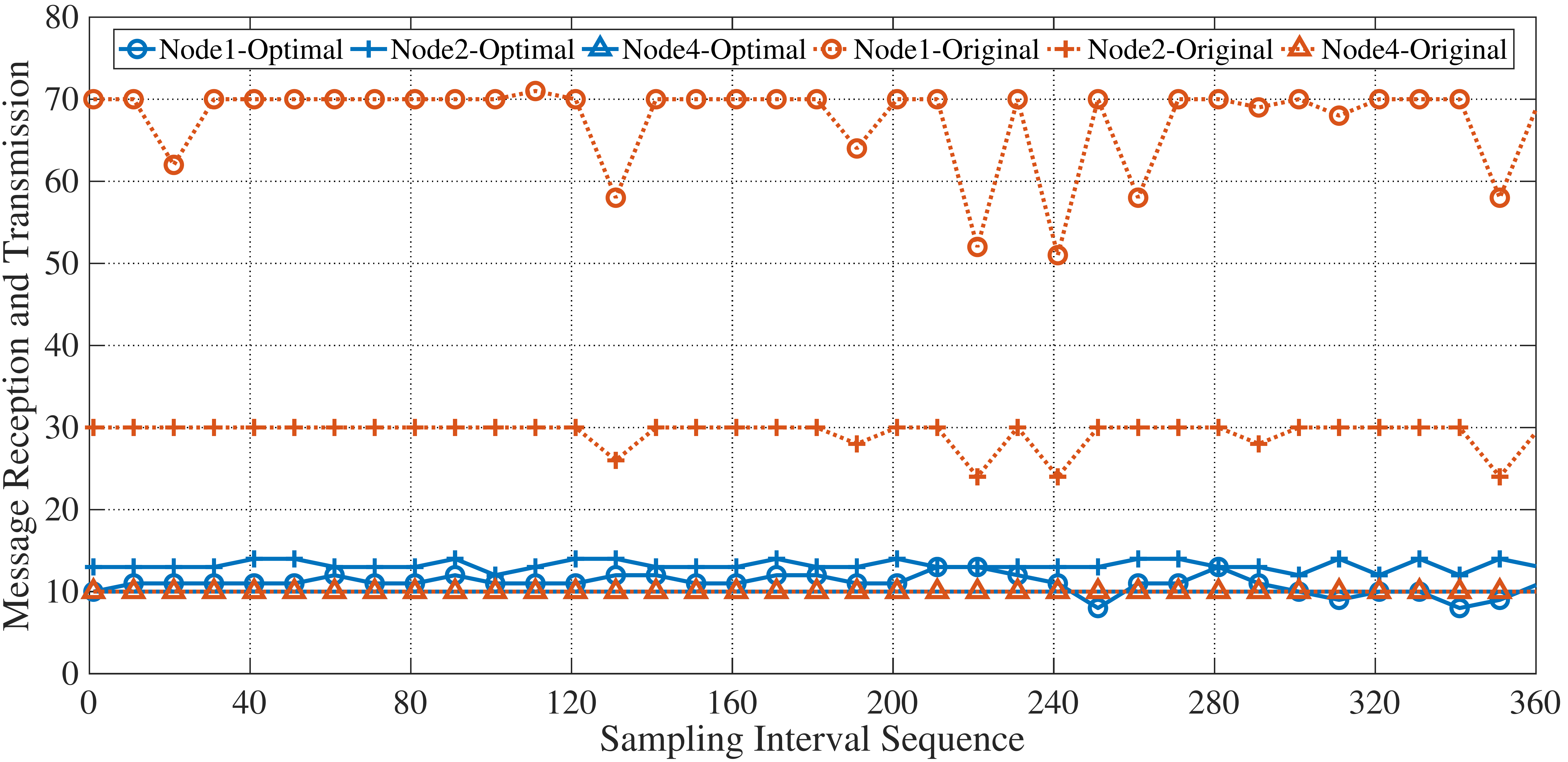}
  \end{center}
  \caption{Number of message receptions and transmissions of sensor nodes with
    and without optimal message bundling for $\SI{3600}{\s}$ with sampling
    interval of $\SI{10}{\s}$.}
  \label{fig:evaluation_message}
\end{figure}
With the novel time synchronization schemes of EE-ASCFR and AHTS, the
synchronization accuracy could be fulfilled as far as the E2E delay of the
bundled message satisfies the synchronization interval: The synchronization 
interval---i.e., E2E delay---exhibited in the evaluation results as shown in
Fig.~\ref{fig:evaluation_onetime} and Fig.~\ref{fig:evaluation_multi}, could
overfulfill the SI requirement (e.g., \SI{10}{\second} SI could lead to 
\SI{2.34}{\us} synchronization accuracy as illustrated in Table.~\ref{tab:time_sync_results_si}), 
which proves that the synchronization accuracy could be strictly followed.

To evaluate the energy efficiency, we indirectly estimate it by comparing the
number of message receptions and transmissions with and without optimal
bundling.  Taking the path of
$0{\leftrightarrow}1{\leftrightarrow}2{\leftrightarrow}4$ as an example, the
number of message receptions and transmissions of the sensor nodes $1,2,4$ in
the experiment of static E2E delay requirement setting in
Section~\ref{sec:evaluation_onetime} is illustrated in
Fig.~\ref{fig:evaluation_message}.
The number of message receptions and transmissions of each sensor node is
counted every \SI{10}{\s} over the period of \SI{3600}{\s}. With the proposed
optimal bundling, all the sensor nodes could maintain their message receptions
and transmissions around the number of $10$. Without the optimal bundling, on
the other hand, the numbers of their message receptions and transmissions are
relatively larger, whose average of $35$ is over triple of that with the optimal
bundling.



\section{Concluding Remarks}
\label{sec:concluding_remarks}
We have proposed an approach to optimize the number of bundled messages at
sensor nodes in a WSN under the constraints of time synchronization accuracy and
E2E delay. To solve the optimal bundling problem, we formulate it as an ILP model
and employ the novel asymmetric time synchronization scheme. To the best of the
authors' knowledge, this is the first work to optimize message bundling for
energy consumption under the joint constraints of synchronization accuracy and
E2E delay in the context of WSNs. The practical evaluation results on a real
testbed demonstrate the long-term and runtime maintenance capability of the
proposed approach.

Note that bundling a larger number of messages results in a longer payload,
which may lead to possible link degradation such as the packet reception ratio
degradation. In this regard, link quality requirements could be introduced to
the optimization model as additional constraints to take into account more
impacts on the overall performance by the bundling procedure.


\bibliographystyle{IEEEtran}%
\bibliography{kks}%

\end{document}